\documentclass[iop]{emulateapj}

\shortauthors{Toft et al.}
\shorttitle{Sub-millimeter galaxies as progenitors of compact quiescent galaxies}
\begin{document}

\title{Sub-millimeter galaxies as progenitors of compact quiescent galaxies}

\author{
S.~Toft \altaffilmark{1}, 
V.~Smol\v{c}i\'{c}\altaffilmark{2}, 
B.~Magnelli\altaffilmark{3}, 
A.~Karim\altaffilmark{3,4}, 
A.~Zirm\altaffilmark{1},
M.~Michalowski\altaffilmark{5,6,18}, 
P.~Capak\altaffilmark{7,8}, 
K.~Sheth\altaffilmark{9},
K.~Schawinski\altaffilmark{10}
J.-K.~Krogager\altaffilmark{1,11}, 
S.~Wuyts\altaffilmark{12},
D.~Sanders\altaffilmark{13}, 
A. W. S.~Man\altaffilmark{1}, 
D.~Lutz\altaffilmark{12},  
J.~Staguhn\altaffilmark{12,14}
S.~ Berta\altaffilmark{12}, 
H.~Mccracken\altaffilmark{15},
J.~Krpan\altaffilmark{2},
D. Riechers\altaffilmark{16,17}
}

\altaffiltext{1}{ Dark Cosmology Centre, Niels Bohr Institute, University of
  Copenhagen, Juliane Mariesvej 30, DK-2100 Copenhagen, Denmark; sune@dark-cosmology.dk}
\altaffiltext{2}{Physics Department, University of Zagreb, Bijeni\v{c}ka cesta 32, 10002 Zagreb, Croatia}
 \altaffiltext{3}{Argelander Institute for Astronomy, Auf dem H\"{u}gel
  71, Bonn, 53121, Germany}
 \altaffiltext{4} {Institute for Computational Cosmology, Durham University, South Road, Durham, DH1 3LE, UK}
\altaffiltext{5} {Scottish Universities Physics Alliance, Institute
  for Astronomy, University of Edinburgh, Royal Observatory, Edinburgh, EH9 3HJ}
 \altaffiltext{6} {Sterrenkundig Observatorium, Universiteit Gent, Krijgslaan 281 S9,
9000 Gent, Belgium}
 \altaffiltext{7}{Spitzer Science Center, 314-6 Caltech, 1201 East California Boulevard, Pasadena, CA 91125, USA}
 \altaffiltext{8}{Department of Astronomy, 249-17 Caltech, 1201 East California Boulevard, Pasadena, CA 91125, USA}
 \altaffiltext{9} {National Radio Astronomy Observatory, 520 Edgemont Road, Charlottesville, VA 22903}
 \altaffiltext{10} {ETH Zurich,Institute for Astronomy, Department of Physics, Wolfgang-Pauli-Strasse 27, 8093 Zurich, Switzerland}
 \altaffiltext{11} {European Southern Observatory, Alonso de Cordova 3107, Vitacura, Casilla 19001, Santiago 19, Chile}
 \altaffiltext{12} {MPE, Postfach 1312, 85741 Garching, Germany}
 \altaffiltext{13} {Institute for Astronomy, 2680 Woodlawn Drive, University of Hawaii,
Honolulu, HI 96822}
 \altaffiltext{14} {The Henry A. Rowland Department of Physics and Astronomy, Johns Hopkins University, 3400 N. Charles Street, Baltimore, MD 21218, USA}
 \altaffiltext{15} {Institut dAstrophysique de Paris, UMR7095 CNRS, Universite Pierre et Marie Curie, 98 bis Boulevard Arago, 75014 Paris, France}
\altaffiltext{16} {Department of Astronomy, Cornell University, 220 Space
Sciences Building, Ithaca, NY 14853, USA}
\altaffiltext{17} {Astronomy Department, California Institute of Technology,
MC 249-17, 1200 East California Boulevard, Pasadena, CA
91125, USA}

\begin{abstract}
Three billion years after the big bang (at redshift $z=2$), half of the
most massive galaxies were already old, quiescent systems with little to no residual
star formation and extremely
compact with stellar mass densities at least an order of
magnitude larger than in low redshift ellipticals, their 
descendants. 
Little is known about how they formed, but their evolved, dense stellar
populations suggest formation within intense, compact starbursts
1-2 Gyr earlier (at $3<z<6$). 
Simulations
show that gas-rich major mergers can give rise to such starbursts which 
produce dense remnants. 
Sub-millimeter selected 
galaxies (SMGs) are prime examples of intense, gas-rich, 
 starbursts. 
With a new, representative spectroscopic sample of
compact quiescent galaxies at $z=2$ and a statistically 
well-understood sample of SMGs, we show that $z=3-6$ SMGs are
consistent with being the progenitors of $z=2$ quiescent
galaxies, matching their formation redshifts and 
their distributions of sizes, stellar
masses and internal velocities. Assuming an evolutionary connection, their space densities also match 
if the mean duty cycle of SMG starbursts is
$42^{+40}_{-29}$ Myr (consistent with independent estimates), which
indicates that the bulk of stars in these massive galaxies were
formed in a major, early surge of star-formation. These results suggests a coherent picture of the formation history of the most massive galaxies in the universe, from their initial burst of violent star-formation through their appearance as high stellar-density galaxy cores and to their ultimate fate as giant ellipticals.
\end{abstract}

\keywords{cosmology: observations -- galaxies: evolution -- galaxies:
  high redshift -- galaxies: starburst -- galaxies: formation-- submillimeter: galaxies}
\vspace{-2cm}
\section{Introduction}
One of the most remarkable discoveries in galaxy evolution studies in
the past years is that up to half of the most massive
galaxies ($\mathrm{log(M_{*}/M_{\odot})>11}$) at $\mathrm{z\approx2}$ are old quiescent systems
with extremely compact structure, corresponding to stellar densities
orders of magnitudes higher than seen in local elliptical galaxies
\citep[e.g.][]{toft2007, vandokkum2008, szomoru2012}. Much effort has
gone into confirming their extreme properties and investigating their
evolutionary path to the local universe. Virial arguments and
simulations indicate that the most important process is
likely to be minor dry merging
\citep[e.g.][]{bezanson2009,oser2012, cimatti2012,toft2012}, but
 observations suggest that other processes are likely also important, e.g. the continuous addition of increasingly larger newly quenched galaxies to the
quenched population with decreasing redshift \citep[e.g.][]{newman2012,carollo2013, cassata2013, krogager2013}.
The formation path of these extreme systems is largely
unknown. Simulations indicate that highly dissipational interactions
on short timescales provide plausible mechanisms for creating compact
stellar populations, either through major mergers
\citep[e.g.][]{naab2007,naab2009}, or dynamical instabilities fed by
cold gas
accretion \citep{dekel2009}. A possible
scenario is major gas-rich mergers at high redshift \citep{wuyts2010},
in which the gas is driven to the center, igniting a massive nuclear
starburst, followed by an AGN/QSO phase
that quenches the star formation, and leaves behind a compact remnant
\citep{sanders1988,hopkins2006, wuyts2010}. This is consistent with
local stellar archaeology studies which imply that massive ellipticals must
have short formation timescales of less than 1 Gyr, \citep[e.g.][]{thomas2005}.

Several authors have pointed out that sub-millimeter galaxies (SMGs)
may be examples of the above scenario
\citep[e.g.][]{blain2004, chapman2005,tacconi2006, toft2007, cimatti2008, capak2008,
  tacconi2008,schinnerer2008,coppin2008,michalowski2010a,smolcic2011},
but see \cite{riechers2013b} for a counter example. The SMG
population is dominated by  galaxies undergoing intense, dust enshrouded
starbursts. A large fraction of SMGs with
measured CO profiles, show double peaked profiles, evident of ongoing
major mergers or rotation \citep[][]{frayer1999, neri2003, sheth2004,
  kneib2005, greve2005, tacconi2006, riechers2011a, ivison2013, fu2013}. 
The auto-correlation length of SMGs 
is similar to that of optically selected QSOs,
suggesting that SMGs and QSOs live in similar mass haloes and that the
ignition of a QSO could be the event that quenches the star formation
in SMGs \citep{hickox2012}. This is consistent with observations suggesting
that the hosts of the most luminous QSOs, i.e those likely associated with the
formation of massive quiescent galaxies, are found to be primarily
major mergers \citep{treister2012, riechers2008}, a result which is cooborated by
\cite{olsen2013} who find that luminous AGN in massive $z\sim2$ galaxies must be triggered by
external processes. Interestingly, \cite{olsen2013} also finds
evidence for low luminosity AGN in the vast majority of massive
quiescent galaxies at $z\sim2$, suggesting that AGN play an active
role in the quenching their star-formation.
The correlation length of SMGs is similar to that of $\mathrm{z\sim2}$ galaxies with
$\mathrm{M_*>5 \times 10^{10}\, M_{\odot}}$ ($\mathrm{r_0=7.66\pm0.78}$),  while $z\mathrm{\sim2}$ galaxies
with  $\mathrm{M_*>10^{11}\, M_{\odot}}$ cluster more strongly
\citep[$\mathrm{r_0=11.49\pm1.26}$][]{wake2011}. 

Recent advances in near infrared (NIR) spectroscopy have made it
possible for the first time to accurately constrain the
age, dust content and past star formation history of the brightest $\mathrm{z\sim2}$
quiescent galaxies through absorption line diagnostics and spectral
fitting in the rest
frame-optical \citep{kriek2009, onodera2010, vandesande2011,
  toft2012, onodera2012, vandesande2012}. These galaxies have spectra
typical of post starburst galaxies, with no detected emission lines,
but with strong Balmer absorption lines, suggesting that they
underwent major starbursts which were quenched 1-2 Gyr prior to the
time of observation (i.e. at $\mathrm{3<z<6}$).
Several of these galaxies show evidence of significant dust abundance
(with $\mathrm{A_V}$'s up to $\sim 1$ magnitude), and they are baryon dominated, as
is the case for local post starburst galaxies \citep{toft2012}.
In combination with their extremely compact
stellar populations these observations suggest that the majority of
the stars in $\mathrm{z\sim2}$ quiescent galaxies
formed in intense, possibly dust enshrouded nuclear starbursts, a
scenario very similar to what is observed in $\mathrm{z\sim2}$ SMGs. 

Velocity dispersions of $\mathrm{z\sim2}$ quiescent galaxies measured
from the width of absorption lines 
are in the range $300-500\, \mathrm{km\, s}^{-1}$
\citep[e.g.][]{toft2012, vandesande2012}, significantly higher than in local
ellipticals of similar stellar mass, but comparable to the FWHM of molecular
lines in $\mathrm{2<z<3}$ SMGs \citep[in the range 
$\mathrm{350-800\, km\, s^{-1}}$,
with a mean equivalent rotational velocity $\mathrm{\left <v_c \right >=
392\pm134 km/s}$;][]{tacconi2006}. The line emitting gas of SMGs, as
traced by high-J CO lines, is
 found to be spatially very compact, with a mean size of 
$\mathrm{\left < R_{e}\right >=2.0\pm0.3 kpc}$ \citep{tacconi2006}, comparable to the mean spatial
extent $\mathrm{\left < R_{e}\right >=1.96\pm0.8 kpc}$, of the stellar
populations in the quiescent $\mathrm{z\sim2}$ galaxies \citep{krogager2013}. We note however that studies of lower-J CO lines suggest
that some SMGs may have more extended CO-disks \citep{ivison2011, riechers2011b}.
The median dynamical masses measured from CO(1-0) for $\mathrm{z\sim2}$ SMGs,  $\mathrm{\left < M_{dyn}\right
>=(2.3 \pm1.4)\times 10^{11}\, M_{\odot}}$ \citep{ivison2011}, is similar to that
measured for $\mathrm{z\sim2}$ quiescent galaxies $\mathrm{\left < M_{dyn}\right>} =(2.5 \pm 1.3)\times 10^{11}\,
\mathrm{M_{\odot}}$ \citep{toft2012}.

Despite the many similarities between SMGs and $\mathrm{z\sim2}$ quiescent
galaxies, a major obstacle in establishing an evolutionary link
between the two galaxy types is their similar redshift
distribution. While the quiescent nature and derived ages for $\mathrm{z\sim2}$
quiescent galaxies suggest they formed at $\mathrm{z\gtrsim3}$ the peak of the known SMG
population was until recently found to be at $\mathrm{z\sim2}$ with very few examples known at
$\mathrm{z\gtrsim 3}$ \citep[e.g.][]{chapman2005},  rendering an evolutionary link between the two
populations unlikely. Recently, improved selection techniques have
however uncovered a substantial tail stretching out to redshifts of
$\mathrm{z\sim6}$ \citep{capak2008, schinnerer2008, daddi2009a,
  daddi2009b, knudsen2010,carilli2010,
  carilli2011,riechers2010,cox2011, combes2012, yun2012,smolcic2012a,
  michalowski2012a, hodge2012,hodge2013, riechers2013}. 

In this paper we present evidence for a direct evolutionary link between
the two extreme galaxy populations, by comparing the properties of two
unique samples in the COSMOS field: (i) a spectroscopically confirmed, representative sample of compact $\mathrm{z\sim2}$ 
quiescent galaxies with high resolution HST/WFC3 imaging, and (ii) a
statistical sample of $\mathrm{z\gtrsim 3}$ SMGs. In Section \ref{section.samples} we
introduce the samples, and in Section \ref{section.results} we present our results. In particular, in Section \ref{section.redshiftdistributions} we show that the
distribution of formation redshifts for the $\mathrm{z\sim2}$
quiescent galaxies is
similar to the observed redshift distribution of $\mathrm{z\gtrsim 3}$
SMGs, and in Section \ref{section.numberdensities} we compare the
co-moving number densities of the two populations.
In Section \ref{section.structure} we derive structural properties of the $\mathrm{z\gtrsim 3}$
SMGs and in Section \ref{section.masssize} show that their stellar mass-size relation is similar to those of $\mathrm{z\sim2}$ quiescent
galaxies.
 In Section \ref{section.dutycycle} and \ref{section.eddington} we
 show that the duty cycle of the $z>3$ SMG starburst (derived assuming
 they are progenitors of $\mathrm{z\sim2}$ quiescent galaxies), is
consistent with independent estimates, and with the formation time
scale derived for $z\sim2$ quiescent
galaxies, assuming they formed in
Eddington limited starbursts. 
 In Section
\ref{section.discussion} we summarize and discuss the results.

Throughout this paper we assume a standard flat universe with
$\mathrm{\Omega_{\lambda}=0.73}$, $\mathrm{\Omega_{m}=0.27}$ and $\mathrm{H_0=71 km
s^{-1} Mpc^{-3}}$. All stellar masses are derived assuming a
\cite{chabrier2003} IMF.

\begin{deluxetable}{llccc}
\tablecolumns{5}

\tablecaption{Sample of $\mathrm{z> 3}$ submillimeter galaxies in COSMOS. The top 11
  galaxies constitute the S/N limited, relatively complete statistical
 sample we use for estimating the comoving number density. The bottom two are
 spectroscopically confirmed $\mathrm{z> 3}$ galaxies which we add to the sample
for structural analysis only. We refer to \cite{smolcic2012a} for
details about the sample. The listed effective radii reported are
circularized, i.e. $\mathrm{r_{e,c}=r_{e,m}\sqrt{b/a}}$, where
$\mathrm{r_{e,m}}$ is the effective radius along the major axis, and
$\mathrm{b/a}$ is the axis ratio. $\mathrm{r_{e,FIR} [kpc]}$ are
restframe FIR sizes from the litterature, measured from high resolution mm observations (see
table notes). For easy comparison to the NIR effective radii, we here quote Gaussian HWHMs.}
\tablehead{
\colhead{} & \colhead{$z$} &  $r_{e,\mathrm{NIR}}$ &\colhead{note}   & \colhead{$\mathrm{r_{e,FIR}}$} \\
                 &                    &        (kpc)              &
                 & (kpc)}
\startdata
AzTEC 1          & $4.64\tablenotemark{a}$                 &   $<2.6$ & unresolved & $1.3-2.7$\tablenotemark{c} \\
AzTEC 3          & $5.299\tablenotemark{a}$               &   $<2.4$&  unresolved &    $<3\pm2$\tablenotemark{d}  \\      
AzTEC 4          & $4.93^{+0.43}_{-1.11}$                          &$<2.5$  & unresolved &-\\
AzTEC 5          & $3.971\tablenotemark{a}$               &   $0.5\pm0.4$&       HST/WFC3 &-\\
AzTEC 8          & $3.179\tablenotemark{a}$               &   $<3.0$         &    unresolved &-\\
AzTEC 10        & $2.79^{+1.86}_{-1.29}$                          &    $0.7\pm0.1$&   -   & - \\
AzTEC 11-S      & $>2.58\tablenotemark{b}$              &   -   & not detected & -  \\
AzTEC 13       &  $> 3.59\tablenotemark{b}$               &    -  & not detected&-\\
AzTEC 14-E     &  $> 3.03\tablenotemark{b}$              &    -     &not detected&-  \\
AzTEC 15       &  $3.17^{+0.29}_{-0.37}$                           &   $5.0\pm0.8$ &    very faint&-\\
 J1000+0234 &  $4.542\tablenotemark{a}$                &  $3.7\pm0.2$     &    -   &-    \\\hline
 Vd-17871     &  $4.622\tablenotemark{a}$               &   $1.3\pm1.4$      &   -    &-  \\
GISMO-AK03  &  $4.757\tablenotemark{a}$               &   $1.6\pm0.6$       &    HST/WFC3 & -
\enddata
\tablenotetext{a}{Spectroscopic Redshift}
\tablenotetext{b}{mm-to-radio flux ratio based redshift}
\tablenotetext{c}{\cite{younger2008}}
\tablenotetext{d}{\cite{riechers2010}}
\label{table.sample}
\end{deluxetable}

\section{Samples}
\label{section.samples}

\begin{deluxetable*}{lclccccrccc}
\tabletypesize{\scriptsize}
\tablecolumns{11}
\tablecaption{Far-infrared SED properties of the $\mathrm{z\gtrsim 3}$ SMG sample.}
\tablewidth{0pt}
\tablehead{
\colhead{         } & 
\colhead{$\mathrm{log (M_{\star})}$\tablenotemark{a}} & 
\colhead{$q_{\rm PAH}$\tablenotemark{b}} &  
\colhead{$\gamma$\tablenotemark{b}} &
\colhead{$U_{\rm min}$\tablenotemark{b}} & 
\colhead{$\mathrm{log(M_{\rm dust})}$ \tablenotemark{b}} & 
\colhead{log($L_{\rm   IR}$)\tablenotemark{b}} &
\colhead{SFR \tablenotemark{b,e}}&
\colhead{$\mathrm{log(M_{gas} )}$\tablenotemark{b}} & 
\colhead{FIR detection\tablenotemark{d}} &  
\colhead{$\mathrm{log(M_{gas} )}_{\mathrm{CO}}$\tablenotemark{f}}\\ & 
\colhead{[$\mathrm{M_{\odot}]}$} & 
\colhead{}& 
\colhead{} &
\colhead{} &
\colhead{$[\mathrm{M_{\odot}}]$} &
\colhead{$[\mathrm{L_{\odot}}]$} &
\colhead{ $\mathrm{[M_{\odot}yr^{-1}]}$}&
\colhead{$[\mathrm{M_{\odot}}]$}&
\colhead{}&
\colhead{$[\mathrm{M_{\odot}}]$}
}
\startdata
AzTEC 1   &$10.9^{+0.1}_{-0.1}$ &  0.47\% & 0.070 &  25.0  &
$9.1\pm0.1$ & $13.36\pm0.09$ &$ 2291\pm 528$&$11.7\pm0.1$&secure & \\
AzTEC 3   & $11.2^{+0.1}_{-0.1}$ &  0.47\% & 0.290 &  25.0  &
$9.3\pm0.1$ &  $13.37\pm0.04$ &$2344\pm  226$&$11.3\pm0.1$&secure   & 10.7 \\      
AzTEC 4   &$11.2^{+0.1}_{-0.1}$&  4.58\% & 0.080 &  25.0  &  $9.6\pm0.2$ &  $13.25\pm0.15$& $1778\pm  733$&$11.6\pm0.2$ &tentative  &\\
AzTEC 5   & $10.9^{+0.5}_{-0.5}$ &  0.47\% & 0.190 &  25.0  &  $9.4\pm0.1$ &$13.43\pm0.02$&$2692\pm  127$ &$11.4\pm0.1$&secure & \\
AzTEC 8   & $11.5^{+0.1}_{-0.1}$&  0.47\% & 0.090 &  25.0  &  $9.7\pm0.1$ &  $13.45\pm0.01$&$2818\pm  66$ &$11.7\pm0.1$ &secure& \\
AzTEC 10  & $10.5^{+0.1}_{-0.1}$&  2.50\% & 0.060 &  5.00   &  $9.6\pm0.1$ &  $12.58\pm0.10$&$380\pm  98$ &$11.8\pm0.1$ &secure  & \\
AzTEC 11-S &  - &  0.47\% & 0.040 &  25.0  &  $9.6\pm0.1$&$13.30\pm0.01$&$1995 \pm46$ &- &secure  & \\
AzTEC 13   & - &  4.58\% & 0.160 &  2.00  &  $9.9\pm0.3$ &  $12.70\pm0.20$&$501 \pm 293$ &- &upper limits&  \\
AzTEC 14-E  & - &  0.47\% & 0.290 &  0.70  &  $9.8\pm0.2$   & $12.48\pm0.18$&$302 \pm 155$ &- &upper limits& \\
AzTEC 15    & $11.2^{+0.1}_{-0.1}$&  3.19\% & 0.010 &  20.0  &  $9.3\pm0.1$&$12.73\pm0.08$& $537\pm 108$&$11.4\pm0.1$ &secure& \\
J1000+0234 & $10.7^{+0.1}_{-0.1}$&  1.77\% & 0.150 &  25.0  &  $9.3\pm0.1$   &$13.17\pm0.09$&$575\pm 275$&$11.8\pm0.4$ &tentative& 10.4 \\\hline
 Vd-17871     &$10.9^{+0.1}_{-0.1}$&  4.58\% & 0.250 &  25.0  &  $9.1\pm0.1$  &$13.09\pm0.06$&$1230 \pm 182$ &$11.2\pm0.2$&secure & \\
GISMO-AK03    &$12.1^{+0.1}_{-0.1}$&  4.58\% & 0.290 &  3.00  &  $9.5\pm0.2$  &$12.66\pm0.19$&$457 \pm 250$ &$11.5\pm0.2$ &secure &
\enddata
\tablenotetext{a}{Derived using {\sc MAGPHYS} from the optical-FIR SED. The errors are
the formal errors associated with the fit, and do not include systematic errors which can be up to $\pm0.5$ Dex , see Section \ref{section.mstar}}.
\tablenotetext{b}{The DL07 model describes the interstellar dust as a mixture of carbonaceous and amorphous silicate
  grains. Here we list the best fitting values of its 4 free parameters:
(i) $q_{{\rm PAH}}$ which controls the fraction of dust mass in the form of PAH grains.
(ii) $\gamma$ which controls the fraction of dust mass exposed to a power-law ($\alpha=2$) radiation field ranging from $U_{{\rm min}}$ to $U_{{\rm max}}$; the rest of the dust mass (i.e., $1-\gamma$) being exposed to a radiation field with a constant intensity $U_{{\rm min}}$.
(iii) $U_{{\rm min}}$ which controls the minimum radiation field seen by the dust ($U_{{\rm max}}$ is fixed to a value of $10^6$).
(iv) $M_{\rm dust}$ which controls the normalization of the SED.}
\tablenotetext{b}{Quantities derived from the best fitting
  DL07 models (see Section \ref{section.fir}).} 
\tablenotetext{d} {``Secure'': The source is relatively isolated and
  detected at $\mathrm{S/N>3}$.  ``Tentative'': The source is detected at $\mathrm{S/N>3}$,
  but the flux density estimates may be affected by bright closeby
  objects.  ``Upper limit'': The source is not detected at $\mathrm{S/N>3}$.  }
\tablenotetext{e}{SFRs are notoriously model dependent, e.g, from
a detailed analysis of all available data for AzTEC-3, and assuming a
top heavy  IMF, \cite{dwek2011} found a significantly lower SFR than
derived here from the DL07 fits.} 
\tablenotetext{f}{Gas masses derived from CO observations
  \citep{schinnerer2008,riechers2010}}
\label{table.mdust}
\end{deluxetable*}

\subsection{Sample of $z\gtrsim 3$ SMGs }
\label{section.z3smg}
Based on dedicated follow-up studies with sub-mm-interferometers (PdBI, SMA, CARMA) and
optical/mm-spectroscopy (with Keck/DEIMOS, EVLA, PdBI) towards
 1.1~mm and $870~\micron$ selected sources  in the COSMOS field
 \cite{smolcic2012a} presented the redshift distribution of SMGs. 
This sample shows a tail of $\mathrm{z\gtrsim 3}$ SMGs, corresponding to a
significantly larger number density at these high redshifts than found
in previous surveys \citep[e.g.][]{chapman2005,wardlow2011, yun2012,
  michalowski2012a}. A possible reason for the difference is that
previous surveys did not have (sub)-mm followup interferometry, and
therefore may be subject to identification
biases. E.g. \cite{hodge2013b} show that many of the galaxies in the
\citet{wardlow2011} sample break up into multiple sources when studied
at high resolution, which enevitably lead to mis-identifications for some of the sources.

Here we use the \cite{smolcic2012a} sample to estimate the comoving number density and other properties
of $\mathrm{z\gtrsim 3}$ SMGs. 
Our starting point is a 1.1~mm-selected sample, drawn from the AzTEC/JCMT 0.15
square degree survey of the COSMOS field (Scott et al.\ 2008), and
observed with the SMA at 890~$\mu$m and $\mathrm{\sim2 \arcsec}$ angular resolution
to unambiguously associate multi-wavelength counterparts \citep{younger2007, younger2009}. The 17 SMGs identified by the SMA follow-up  form a
statistical sample as they are drawn from a signal-to-noise limited
($\mathrm{S/N_{1.1mm}}>4.5$), and flux-limited
($F_\mathrm{1.1mm}\gtrsim4.2$~mJy), 1.1~mm-selected sample, drawn from
a contiguous
area of 0.15 square degrees. We include one more SMG in this sample,
J1000+0234 ($F_\mathrm{1.1mm}=4.8\pm1.5$~mJy,
$\mathrm{S/N_{1.1mm}}\sim3$),  which is confirmed to be at $\mathrm{z=4.542}$
\citep{capak2008, schinnerer2008,schinnerer2009}. Nine out of these 18 interferometrically detected galaxies have spectroscopic redshifts
(four are confirmed to be at $\mathrm{z\gtrsim 3}$; \citealt{capak2009,
  capak2010,schinnerer2009,riechers2010}; Karim et al., in prep), while for the remainder precise
photometric redshifts ($\sigma_\mathrm{{\Delta z / (1+z_\mathrm{spec})}}=0.09$)
were computed by \cite{smolcic2012a}.  
The $\mathrm{z\gtrsim 3}$ SMGs from this sample are listed in Table \ref{table.sample}. 
The top 11 objects constitute our statistical sample. We will use these
in the following sections  to estimate the redshift distribution and
comoving number density of $z\gtrsim 3$ SMGs. The bottom two objects, are
additional spectroscopically confirmed $\mathrm{z\gtrsim 3}$ SMGs in the COSMOS
field which we add to the sample for structural analysis only. 

The flux-limited sub-mm selection ensures a relatively homogenous  
sample of the most intensely starforming  dust obscured galaxies at ${z\gtrsim3}$: Due
to the negative k-correction, the sub-mm flux detection limit
corresponds roughly to a cut in SFR over the considered redshift
range. 
 Note that while a fraction of single dish detected SMGs break up into
multiple components when studied with interferometry at $\le
2\arcsec$ resolution, this is only the
case for two of the galaxies studied here (AzTEC 11 and 14). In the
present study we assume that the close individual components are
related and count them as one in the number density calculations
 (thus assuming they will eventually merge
into one galaxy). As the galaxies are not resolved in the MIR-mm
photometry, derived properties (infrared luminosities, star formation
rates, dust masses etc) are for the combined systems. Neither of the
two galaxies are detected in the optical-NIR so the derived sizes and stellar
masses for the sample are not affected.  



\subsection{Far-infrared emission of the $\mathrm{z\gtrsim 3}$ SMGs}
\label{section.fir}
In order to directly constrain the SFRs, dust and gas masses of the $\mathrm{z\gtrsim 3}$ SMGs, we made use of 
the (sub)-mm (AzTEC, LABOCA, MAMBO, SMA, CARMA, PdBI) and  FIR (Spitzer MIPS, Herschel PACS and SPIRE)
observations of the COSMOS field \citep{sanders2007,lutz2011,
  oliver2012,scott2008,aretxaga2011,bertoldi2007,younger2007,younger2008,smolcic2012a,smolcic2012b}. 
The Herschel data consist of deep PACS 100 and 160$\,\mu$m observations, taken as
part of the PACS Evolutionary Probe
\citep[PEP,][]{lutz2011}
guaranteed time key programme, and SPIRE 250, 350 and 500$\,\mu$m
observations taken as part of the \textit{Herschel} Multi-tiered
Extragalactic Survey \citep[HerMES, ][]{oliver2012}.

PACS and SPIRE flux densities were measured using a PSF fitting analysis \citep{magnelli2009,roseboom2010}, guided by the position of sources detected in the deep COSMOS 24 $\,\mu$m observations from the Multiband Imaging Photometer \citep[MIPS;][]{rieke2004} onboard the \textit{Spitzer} Space Observatory \citep[3$\sigma$$\,\thicksim\,$45$\,\mu$Jy;][]{lefloch2009}.
We cross-matched our $z\gtrsim 3$ SMGs sample with this MIPS-PACS-SPIRE catalogue using a matching radius of 2\arcsec.
Results of these matches were all visually checked.
For $z\gtrsim3$ SMGs not included in the MIPS-PACS-SPIRE catalogue
because of a lack of MIPS counterpart, we compute their PACS and SPIRE
flux densities using a PSF fitting analysis guided by their
positions. Further details of the FIR photometry are presented in Smol\v{c}i\'{c} et al.\ (in prep.). 
%

Among the 13 $\mathrm{z\gtrsim 3}$ SMGs, 9 have secure mid/far-infrared detections, 2 have tentative mid/far-infrared detections and 2 are undetected at infrared wavelengths. 
From the FIR--mm SED 
of the $\mathrm{z\gtrsim 3}$ SMGs, we infer their infrared luminosities
and dust masses using the dust model of \citet[][hereafter
DL07]{draine2007} as described in detail in \cite{magnelli2012}.
The infrared luminosity ($L_{{\rm IR}}$) is derived by integrating
the best fitting normalized SED templates from the DL07 library from rest-frame 8
to 1000$\,\mu$m. From these we can accurately estimate the star-formation
activity of the $\mathrm{z\gtrsim 3}$ SMGs, using the standard $L_{\rm IR}$-to-SFR conversion of \citep{kennicutt1998}, assuming a Chabrier IMF :
\begin{equation}
\label{equation.sfr}
{\rm SFR\,[M_{\odot}\,yr^{-1}]\,=\,10^{-10}\,}L_{{\rm IR}}\,{\rm [L_{\odot}]}.
\end{equation}

Finally we estimate the gas masses of the sample through
$\mathrm{log(M_{gas}/M_{dust})=-0.85*Z+9.4}$ \citep{leroy2011}, where
$\mathrm{Z=2.18\times
  log(M_{\star})}$$\mathrm{-0.0896*log(M_{\star})^2-4.51}$ \citep{erb2006,
  genzel2012} \footnote{thus making the assumption that the
  mass-metallicity relation at $z\sim2$ apply to galaxies at $z>3$}. This method has been used successfully in the local
Universe \citep[e.g.][]{leroy2011,bolatto2011} as well as at high
redshift \citep{magdis2011, magdis2012, magnelli2012}. Assumptions and
limitations of this method in the case of high redshift galaxies are
extensively discussed in \cite{magnelli2012}. 
Results of the FIR SED fits and derived quantities are summarized in
Table \ref{table.mdust}, and used in the following analysis to
establish an evolutionary link between ${z\gtrsim 3}$ SMGs and quiescent
${z\sim2}$ galaxies. 
 The derived gas masses are comparable to or larger than the derived
 stellar masses: $\mathrm{\left< f_g
   \right > = \left< M_{gas}/(M_{\star}+M_{gas})\right>=0.71\pm0.03}$, in
 agreement with the high gas fractions found in previous studies of
 high redshift SMGs \citep[e.g.][]{carilli2010,
   riechers2011b}.  We caution however, that gas masses estimated from FIR
 SED fits are relatively uncertain (potentially up to a factor of 
 5 -10).  E.g. in Table \ref{table.mdust} we list gas masses for two objects in our sample
which have independent estimates derived from CO line emission. These
are significantly different from our SED estimates. The main factors
contributing to the uncertainties in the SED estimates are that the
sub-mm measurements dont trace cold gas very well, in which the
(sub)mm/CO flux ratio is much lower than in the starburst nucleus, but
where a lot of the gas mass may reside. The others are the metallicity
correction (which has a large scatter) and the assumption about the gas-to-dust ratio.
The main factors contributing to the uncertainty of the CO measurements is the assumed $\alpha_{\mathrm{CO}}$ which can be uncertain by a factor $>2$, and the excitation corrections, which can be uncertain by a factor of $\sim 2-4$.
\subsection{Stellar Mass Estimates for the $z\gtrsim 3$ SMGs}
\label{section.mstar}
We estimate stellar masses of the $z\gtrsim 3$ SMGs from their UV-MIR ($8\micron$) broad band photometry as described in Smol\v{c}i\'{c} et
al.\ (in prep.). Briefly, stellar masses were derived by fitting the
observed broadband UV-MIR spectral energy distributions with the {\sc
  MAGPHYS} code \citep{dacunha2008}. The stellar component in the {\sc
  MAGPHYS} models is based on  \cite{bc03} stellar population
synthesis models assuming various star formation histories
(exponentially-declining SFHs (with random timescales) + superimposed
stochastic bursts)  and a \cite{chabrier2003} IMF. The stellar masses for the
SMGs, and their formal uncertainties drawn from the probability
distribution function (generated from the $\chi^2$ fit values by
{\sc MAGPHYS}) are given in Tab.~\ref{table.mdust}. We note however,
that stellar masses for SMGs are strongly dependent on the assumed
star formation histories, and may lead to systematic discrepancies of
$\pm0.5$~dex given different assumptions and stellar population
synthesis models \citep[see Table.~1 in][]{michalowski2012b}, and
whether or not emission lines are included in the templates \citep{schaerer2013}. 
For example, using the double SFHs implemented in {\sc GRASIL}
\citep{silva1998,iglesias2007}, we find systematically higher stellar
masses, consistent with the results from \cite{michalowski2010b}.
On the other hand, dynamical mass considerations based on CO line
observations for two objects in our sample \citep{schinnerer2008,
  riechers2010} suggest lower stellar masses than inferred by {\sc
  MAGPHYS}. Hence, here we adopt the middle values, i.e.\ the stellar
masses computed by {\sc MAGPHYS}+BC03, noting that these may be
subject to systematic uncertainties.


\subsection{Sample of ${z\sim2}$ Compact Quiescent Galaxies}
\label{section.quiescent}

It is well established from deep multi-waveband photometric surveys,
that a substantial population of quiescent
massive galaxies with extremely compact structure exists at ${z\sim2}$
\citep{daddi2005, toft2005, trujillo2006,
  toft2007,franx2008,toft2009, williams2010,
  vandokkum2010,brammer2011,szomoru2011,damjanov2011, szomoru2012, newman2012, cassata2013}.
Samples of spectroscopically confirmed, ${z\sim2}$ quiescent galaxies  with
accurate stellar population model fits and high angular resolution space based
NIR imaging are much more sparse \citep{vandokkum2008}. As a high quality comparison
set to the $z> 3$ SMGs we use the sample of \citet[][K13 hereafter]{krogager2013}. This sample consist of 16 spectroscopically confirmed massive quiescent galaxies, selected from the 3DHST survey in the COSMOS field by requiring 
strong $4000${\AA} breaks in the grism observations. As shown in
K13 this effectively selects a representative sample of massive (logM$>$10.9) evolved quiescent galaxies
around ${z\sim2}$. The high S/N grism spectra around the break in combination with multi-waveband   photometry from the
COSMOS survey allows for strong constraints on the stellar populations
including stellar masses, dust contents, mean stellar ages, i.e the
time elapsed since the last major episode of star formation, as well as
formation redshifts (derived from the stellar ages).  The sample is
also covered by high resolution NIR imaging with HST/WFC3 from the
CANDELS survey \citep{grogin2011,koekemoer2011} yielding accurate
constraints on the rest-frame optical surface brightness profiles, and effective radii
($\mathrm{r_e}$).

\section{Results}
\label{section.results}

\subsection{Redshift Distributions}
\label{section.redshiftdistributions}

\begin{figure}[ht]
\includegraphics[ scale=0.43,angle=0]{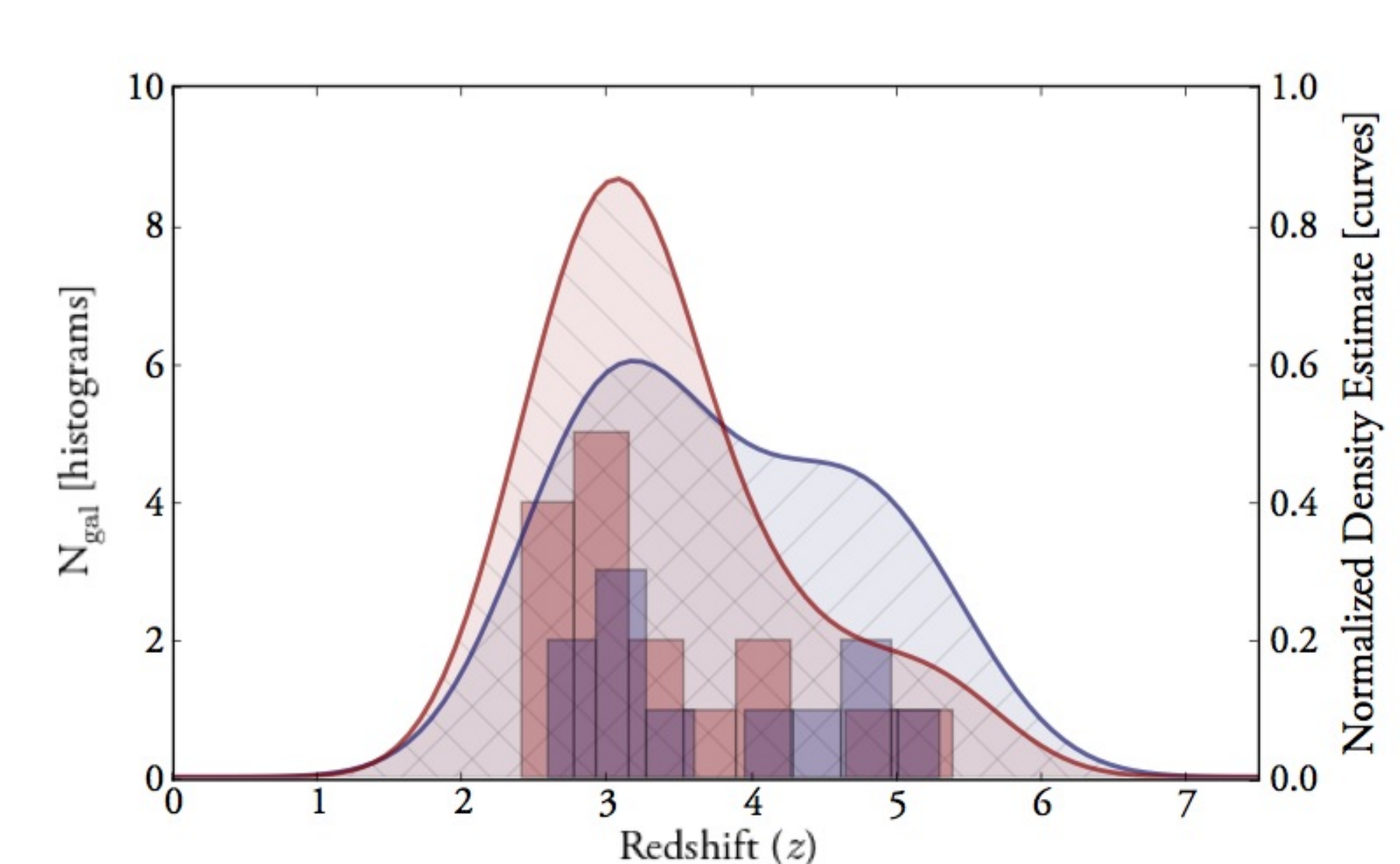}
\caption{Comparison of the redshift distribution of $z\gtrsim3$ SMGs and the
  formation redshifts of $z\sim2$ quiescent galaxies.
 The red histogram shows the distribution of formation redshifts estimated for
  a spectroscopically confirmed sample of compact quiescent galaxies
  at ${z\sim2}$, from their observed  redshift and derived luminosity
  weighted ages of their stellar populations (K13). 
The blue histogram shows the distribution of redshifts of the statistical sample of ${z\gtrsim3}$ SMGs. 
Note that the galaxies which only have
  lower limits on the redshifts have been placed in the bins
  corresponding to those limits, and that the histogram includes two galaxies
  where the best fitting redshift is slightly below 3, but for which a
  $z>3$ photometric redshift solution falls within the $99\%$ confidence interval. 
The smooth curves shows probability density distributions (kernel
density estimates (KDEs)) of the two populations.
 The two redshift distributions are similar, consistent with the
 hypothesis that ${z\gtrsim 3}$ SMGs are the direct progenitors of ${z\sim2}$ cQG.}
\label{figure.zform}
\end{figure}

From the spectroscopic redshifts and mean stellar ages available for the quiescent ${z\sim2}$
galaxy sample described in Section \ref{section.quiescent}  we can estimate
the distribution of their formation redshifts. In Figure
\ref{figure.zform} this distribution is compared to the observed redshift distribution
of the sample of ${z\gtrsim 3}$ SMGs described in Section \ref{section.z3smg}. 
Due to the small number of galaxies  in both samples, a one to one
correspondence is not expected. However, we stress that the two
distributions are similar, with a peak at ${z\sim3}$ and a tail
towards higher redshifts.  
A two-sample Kolmogorov-Smirnov (K-S) test yields a statistic of 0.29
with a p-value of 54\%, and is thus not inconsistent with the
two redshift distributions being drawn from the same parent distribution.

\subsection{Co-moving Number Densities}
\label{section.numberdensities}
The next step in establishing an evolutionary connection between
${z\gtrsim 3}$ SMGs and quiescent galaxies at ${z\sim2}$ is to compare their co-moving
number densities.
The comoving number density of massive quiescent galaxies as a function of redshift is well
constrained from photometric redshift and stellar population model
fits to deep multi-waveband photometry
\citep[e.g.][]{williams2010,brammer2011}.  \cite{brammer2011} studied the
number density evolution of star forming and quiescent galaxies
(separated using their rest-frame UVJ colors) in a
sample complete at stellar masses $\mathrm{log(M/ M_{\odot})>11}$ out to
${z\sim2.2}$. Here we
estimate a comoving number density of $\mathrm{6.0 \pm 2.1 \times 10^{-5} Mpc^{-3}}$ for 
quiescent galaxies at ${z\sim2}$ with 
$\mathrm{log(M/ M_{\odot})>11}$ as the mean
of the densities measured at ${z=1.9}$ and ${z=2.1}$ by
\citet{brammer2011}. The error includes a contribution of cosmic
variance of $12\%$ \citep{moster2011}.
  
To derive the surface number density of ${z\gtrsim 3}$ SMGs we take all SMGs
from the 1.1~mm-selected COSMOS sample that could lie at ${z> 3}$ given
their lower or upper 99\% confidence levels of the photometric
redshift 
\citep[reported in Tab.~4 in][]{smolcic2012a}. We then derive an
average value of the surface density by taking the most probable
photometric redshift (or spectroscopic redshift where
available)\footnote{Taking the most probably photometric redshift
  yields that 9 SMGs (AzTEC1, AzTEC3, AzTEC4, AzTEC5, AzTEC8, AzTEC13,
  AzTEC14E, AzTEC15 \& J1000+0234), are at z$>$3}, and the
lower\footnote{In this case 8 SMGs (AzTEC1, AzTEC3, AzTEC4, AzTEC5,
  AzTEC8, AzTEC13, AzTEC14E \& J1000+0234), are at z$>$3} and
upper\footnote{In this case 10 SMGs (AzTEC1, AzTEC3, AzTEC4, AzTEC5,
  AzTEC8, AzTEC10, AzTEC11S, AzTEC13, AzTEC14E \& J1000+0234), are at
  z$>$3} surface density values by taking the limiting redshifts
corresponding to the 99\% confidence intervals of the photometric
redshifts. This yields a surface density of $z\gtrsim 3$, bright ($F_\mathrm{1.1mm}\gtrsim4.2$~mJy) SMGs of $60\pm10$~deg$^{-2}$. Note that conservatively excluding from the analysis all 3 SMGs in the sample that are not significantly detected at other wavelengths (AzTEC11S, AzTEC13, AzTEC14E), and thus only have lower redshift limits, yields a surface density of 40~deg$^{-2}$. 

The derived surface density values for ${z\gtrsim 3}$ SMGs may be
subject to systematic effects. The completeness of the AzTEC/JCMT
COSMOS survey, shown in Fig. 8 in \citet{scott2008}, is roughly
50\%, 70\%, and 90\% at  $F_\mathrm{1.1mm}=4.2, 5$, and
$6$~mJy. Taking this into account in combination with the deboosted
1.1~mm fluxes of the SMGs \citep[see][]{younger2007, younger2009}
yields that the derived surface densities could be roughly a factor of
1.5 higher than that reported above. On the other hand, the AzTEC/JCMT
COSMOS field may be overdense \citep[][]{austermann2009}, which would
imply that the true ${z \gtrsim 3}$ SMG surface density averaged over larger area would be lower.

Our best estimate of the co-moving number density of ${3<z<6}$ SMGs
is $\mathrm{2.1 \pm 0.4\times 10^{-6}~Mpc^{-3}}$, significantly lower
than the space density of ${z\sim2}$ quiescent galaxies. This is expected
as  ${z\gtrsim 3}$ galaxies only enter the mm-selection criterion
during their intense starburst phase. In Section
\ref{section.dutycycle} we use the observed difference in co-moving
number densities to constrain the duty cycle of the SMG starbursts.

\begin{figure*}
\includegraphics[scale=0.9,angle=0]{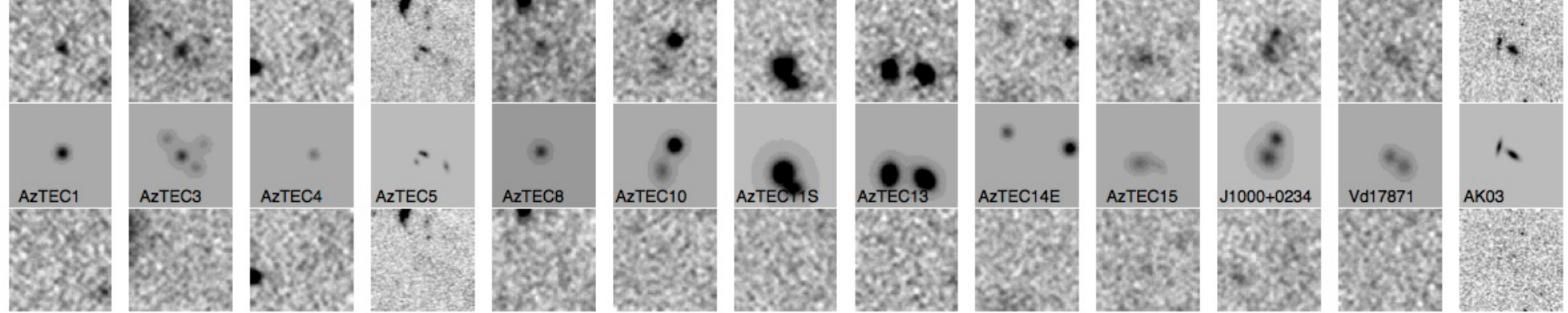}
\caption{{Gallery of ${z> 3}$ SMGs. Top: 
  NIR images, 8'' on a side. For AzTEC5 and GISMO-AK03 we show
  HST/WFC3 F160W images from the CANDELS survey. For the rest we show stacked
Y, J, H and K band images from the UltraVISTA survey. {Middle:}
Sersic $n=1$ {\sc galfit} models of the 2D surface brightness distributions of the
SMGs and their nearby companions. {Bottom:} Residual images,
i.e.  the original images in the top panel, subtracted the best
fitting models in the middle panel.}  
}
\label{figure.gallery}
\end{figure*}

\subsection{Rest-frame UV-Optical Structure of z$>$3 SMGs: Disks,
  Spheroids or Mergers?}
\label{section.structure}
The high redshifts and large amounts of dust in the ${z\gtrsim 3}$ SMGs renders
them extremely faint in the rest-frame UV and optical, despite their
high stellar masses and star formation rates. This makes it challenging to
constrain their structure. To achieve the least biased estimates of
the distribution of stellar mass one would need to study the surface
brightness distributions in the rest-frame optical/NIR, or as close to these wavebands as
possible. Ideally, the observations would be done in the observed mid infrared,
but at the low spatial resolution of current facilities (e.g.\ Spitzer)  the
galaxies remain unresolved. Until JWST becomes operational the best that can be
achieved is to
study the galaxies in the observed NIR. For most of the galaxies in the
sample this wavelength range  probes rest-frame wavelengths around the $4000$~{\AA} break, and
thus should be a relatively good tracer of the stellar mass
distribution. For two galaxies (AzTEC5 and GISMO-AK03) we use space based NIR
imaging with HST/WFC3, which is available from the CANDELS survey. This is preferable to groundbased imaging given 
the higher resolution (FWHM$\sim0.2\arcsec$). For the remaining galaxies we use deep NIR
imaging provided by the UltraVista survey
(5$\sigma$ AB depths range from 23.7 in the K-band to 24.6 in the
Y-band, McCracken et al 2012).
The resolution of these observations is lower (FWHM$\sim$0.8\arcsec), but it
has been demonstrated that relatively unbiased sizes (down to a
fraction of the FWHM$_{PSF}$) can be derived
from such data when the S/N is high and the PSF is well known
\citep[e.g.][]{trujillo2006,toft2009,williams2010}. To increase the
S/N we stack the Y, J, H and K band images. 

Postage stamp images of the galaxies are shown in the top panel of
Figure \ref{figure.gallery}. NIR counterparts of 10 of the 13 sources
are detected, 8 of which have relatively high signal to noise (the faintest ones
have S/N$\sim10$). 
We fit 2D Sersic models to their surface brightness distributions with {\sc{galfit}}
\citep{peng2002}, using similarly stacked images of nearby stars as PSF
models. 
We find the sersic n to be relatively poorly constrained from the data. Leaving it free in the fits in all cases results in low values
$n <2$, with a median value of $\mathrm{\left<n\right>=0.6\pm0.1}$,
but with relatively large errors. To limit the degrees of freedom in
the fits we therefore fix it to $n=1$. The reduced $\chi^2$ of these fits
are in all cases similar to those with n free, and better than fits
with n fixed to 4.
 

The best fitting effective radius encompassing half the light of the model, are
reported in Table \ref{table.sample}. Half of the detected galaxies
(5) have close companions. In these cases we model both components
simultaneously and report the parameters for the main component
(closest to the center of the mm-emission). Also listed in Table
\ref{table.sample} are restframe FIR sizes for two galaxies in our
sample derived from interferometric sub-mm imaging observations. These
agree with the sizes derived from the NIR data. The
restframe FIR sizes directly measures the extend of the
starforming regions which we hypothesize evolves into the compact
stellar populations at z=2, so the agreement is encouraging. 

Our analysis shows, that apart from being very compact, the $z> 3$ SMGs 
are not isolated, smooth single component galaxies. All the detected galaxies show evidence of close companions or clumpy
sub-structure (see Figure \ref{figure.gallery}). From these observations alone, it is not possible to deduce
whether this is due to chance projections, ongoing minor/major mergers, or perhaps multiple
star forming regions in individual galaxies, as
resolved photometry and spectroscopy is not available.  We note however that the
two galaxies with HST/WFC3 data appear to have well separated
individual components of comparable brightness, favoring the merger
interpretation. This is consistent with direct observational evidence
for SMGs being major mergers, i.e. having multiple close
components at the same redshift \citep[e.g.][]{fu2013,ivison2013}.
Simulations suggest that the timescale for major mergers
are typically $0.39\pm0.30$  Gyr \citep{lotz2010}. The cosmic time available between the observed epoch of the SMGs at
z=3-6 and their proposed remnants at z=2 is 1-2 Gyr. If (some of) the
SMGs are major mergers, there is thus sufficient time available for
them to coalesce to a  single quiescent remnant 
at $z=2$. 

In the local universe most star forming
galaxies are well fit by exponential disk profiles corresponding to
n=1 \citep{wuyts2011}, while irregular galaxies and  (pre-coalescence) mergers
 are often best fit by models with lower n-values (n$<$1). At the S/N
and resolution of the galaxies in the Ultravista data the confidence in derived Sersic
parameters are limited. However, the persisting low values found for
the whole $z> 3$ SMG sample, including the two galaxies with the higher resolution HST/WFC3 data, suggests
that the galaxies are more consistent with disks or mergers than spheroids. 
A similar conclusion was found for a 
sample of 22 SMGs at $1<z<3$ with HST/WFC3 data \citep{targett2012}
for which the
majority were best fit by low n Sersic models, with a mean
$\mathrm{\left< n \right>=1.2 \pm 0.1}$.  
If $z> 3$ SMGs are
progenitors of $z\sim2$ quiescent galaxies, then their evolution must
include a transformation of their surface brightness profiles which
increase their Sersic indices, as surface brightness profiles of
quiescent galaxies at ${z\sim2}$ are more centrally
concentrated \citep{wuyts2011,szomoru2011,bell2012},  e.g. the sample of K13 
 has $\mathrm{\left < n \right >=4.0 \pm0.1}$.   We discuss 
a possible mechanism for this transformation in Section \ref{section.discussion}.

\subsection{Mass-Size relation}
\label{section.masssize}
\begin{figure}
\includegraphics[scale=0.45,angle=0]{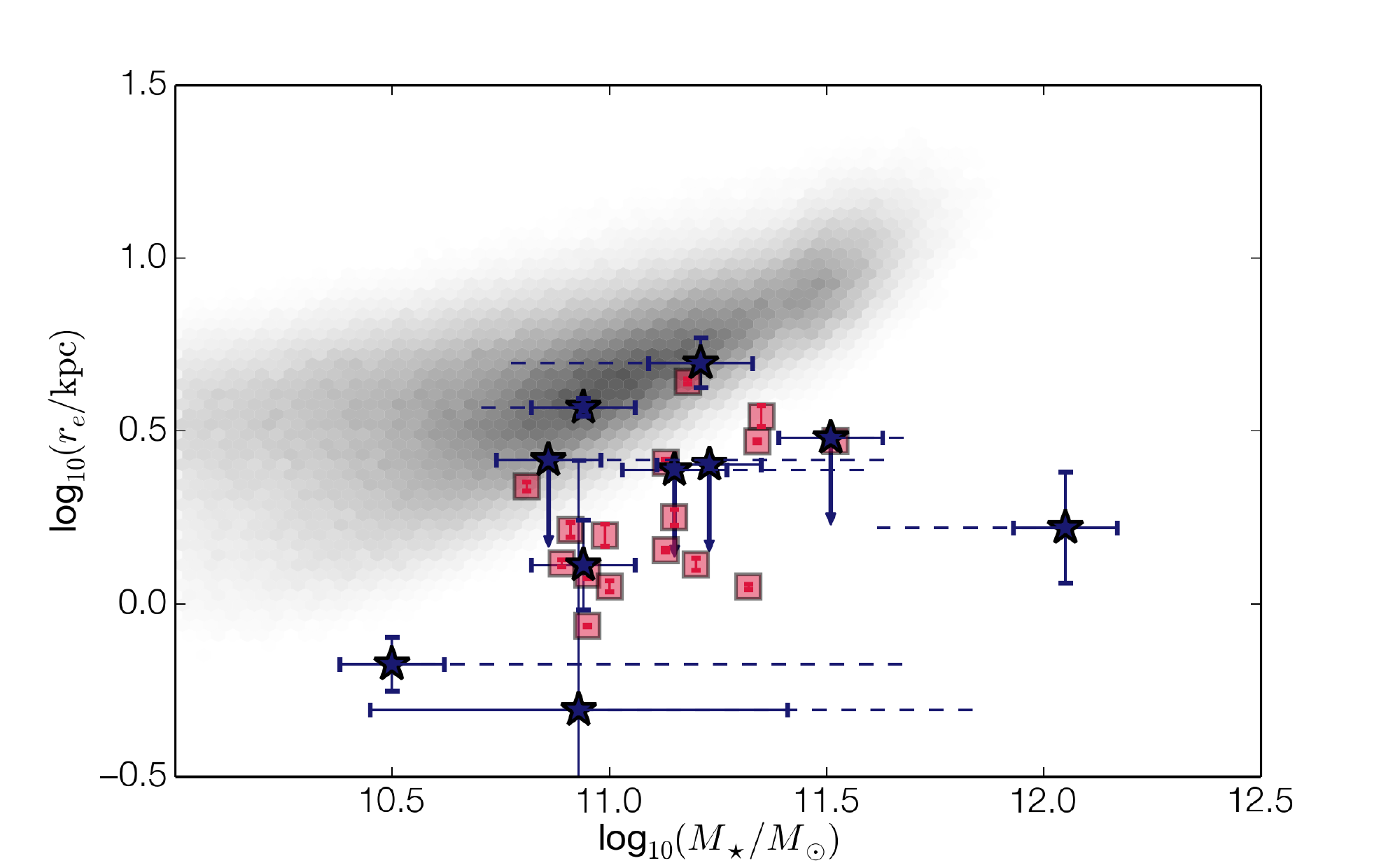}
\caption{{Comparison of the stellar mass-size plane of $z\gtrsim3$ SMGs, $z\sim2$
   quiescent galaxies and local galaxies.}
The red points represent ${z\sim2}$ quiescent galaxies. Black points represent
$z\gtrsim3$ SMGs. For the latter, the solid error bars represent the errors
associated with the {\sc MAGPHYS} SED fits. The dotted error bars are possible
systematic errors that extend to values we derive using the
\cite{michalowski2010a} templates.
The grey cloud shows the mass-size distribution of massive local galaxies in the
SDSS survey.
The mass-size distributions of SMGs is similar to that of $z\sim2$
quiescent galaxies, significantly offset from the local relation, consistent with a direct
evolutionary connection between the two populations. 
}
\label{figure.masssize}
\end{figure}

Combining the derived stellar masses and effective radii of the $3<z<6$ SMGs, in
 Figure \ref{figure.masssize} we compare their stellar mass-size distribution 
to that of $z=2$ quiescent galaxies and of massive early type  galaxies in the local universe. 
Two of the ten NIR-detected SMGs are relatively extended with
effective radii comparable to those in local galaxies of similar mass. Both of these (AzTEC 10 and AzTEC 15) appear from their NIR images to be ongoing  mergers. 
The remaining 8 galaxies are 
extremely compact, with $\mathrm{r_e \lesssim 2.5~kpc}$. Four are unresolved in the Ultravista data.  For these we adopt upper limits on their effective radii corresponding to $\mathrm{0.5\times FWHM_{PSF}}$. 

The stellar mass-size distribution of the $3<z<6$ SMGs is
similar to that of ${z\sim2}$ quiescent galaxies. Both populations are smaller by an average factor of $\sim 3$ than local galaxies of similar mass.
From the derived quantities we can infer the mean internal stellar mass surface
densities within the effective radius
($\mathrm{\Sigma=0.5M_{\star}/\pi r_{e}^2}$) of the $z>3$ SMGs and $z=2$
compact quiescent galaxies (cQGs) which we find to be
similar  $\mathrm{\left<log(\Sigma)\right >_{SMGs} \sim9.9 \pm 0.1
  M_{\odot}{kpc}^{-2}}$, $\mathrm{\left<log(\Sigma)\right >_{cQGs} \sim9.8 \pm 0.1 M_{\odot}{kpc}^{-2}}$  more than an order of magnitude higher than in
local early type galaxies of similar mass. This is consistent with
a picture where the SMGs passively evolve into compact quiescent galaxies after their 
starbursts are quenched.

\subsection{Duty Cycle of SMG Starbursts}
\label{section.dutycycle}
The observed
space density of  $z\gtrsim 3$ SMGs is a factor of $\mathrm{\sim30}$ lower than the space
density of ${z\sim2}$ quiescent galaxies (see Section \ref{section.numberdensities}). However, the SMGs only enter the sub-mm selected ($F_\mathrm{1.1mm}\gtrsim4.2$~mJy)
sample during their intense starburst phase where they have very high
star formation rates. The duration of this phase, i.e.\ the duty cycle
  $\mathrm{t_{burst}}$, which ends when the supply of gas is depleted or the
  star formation is quenched, e.g. by feedback from supernovae or
  active galactic nuclei, has been estimated to be in the range 40-200
  Myr, based on gas depletion timescales
  \citep[][]{greve2005,tacconi2006, riechers2011c} 
and clustering analysis \citep{hickox2012}. 
If we assume that all the ${z\gtrsim 3}$ SMGs evolve into  ${z\sim2}$ quiescent
galaxies, and that they only undergo one SMG phase we can
estimate the average duty cycle of their starbursts from the observed comoving number
densities of the two populations, as
\begin{equation}
\mathrm{t_{burst}=t_{obs}\times (n_{SMG,z\gtrsim 3}/n_{q,z=2})},
\end{equation}
where $\mathrm{t_{obs}}$ is the cosmic epoch corresponding to the redshift
interval ${3<z<6}$ from which the ${z\gtrsim 3}$ SMGs are selected.
Using the comoving number densities we can
thus constrain the  duty cycle of the SMGs to
$t_{burst}=42^{+33}_{-15}$Myr. 
This number however does not include possible
systematics uncertainties on the number density of SMGs discussed in Section
\ref{section.numberdensities}. 
If we conservatively assume the two extreme cases where (i) the SMG sample is
 100\% complete, and the field is three times overdense, and (ii)
the sample is a factor of 1.5 incomplete but not overdense, the derived timescale
are in the range $\mathrm{14<t_{SMG}<62\, Myr}$. The systematic uncertainties on
the timescale is thus of the order of 24~Myr. Therefore our constraints on 
  the average dutycycle in ${z\gtrsim 3}$ SMGs is
 $\mathrm{t_{burst}=42_{-29}^{+40}Myr}$, where the errors have
 been added in quadrature. This value is consistent with the
 independently estimated duty cycles based on gas depletion time
 scales, thus affirming the idea that ${z\gtrsim 3}$ SMGs are progenitors
 of ${z\sim2}$ quiescent galaxies.  
The derived timescale does not depend strongly on the
 ${z=6}$ upper limit adopted for the SMG redshift distribution. Adopting 
  limits of z=5.5 or z=7 instead, leads to timescales of 44 and 37
  Myr, respectively. 
We note that the validity of the timescale calculation presented here, relies on the assumption of a
direct evolutionary connection between the two populations, implying
  that all $z=2$ quiescent galaxies were once $z>3$ SMGs, and all
  $z>3$ SMGs evolve into $z=2$ quiescent galaxies.

\subsection{Star formation Rate and Timescale of z=2 Quiescent
  Galaxies During their Formation }
\label{section.eddington}
We can infer a lower limit on the star formation rate of the z=2 quiescent galaxies during their formation by assuming that they started forming stars at $z=10$ and did so at a 
constant rate until their inferred formation redshifts.  The minimum average SFR needed 
to acquire their observed stellar masses at z=2 calculated in this way
is 
$\langle\mbox{SFR}_{\rm min}\rangle = 115 \pm 5\, M_\odot\,\mbox{yr}^{-1}$.
This is a factor of $>3$ larger than the observed average SFR in 
star-forming Lyman break galaxies (LBGs) at $z>3$ \citep{carilli2008}.
Furthermore, the space density of  $z\sim2$ quiescent galaxies with
logM$_{*}$/M$_{\odot}>11$, is 5, 10 and $>100$ times larger than that of
similar mass LBGs at z=4,5 and 6 respectively \citep{stark2009}.
Their progenitors must therefore have had much larger SFRs and are missing from 
LBG samples. This suggests that they must be dust obscured starburst
galaxies.

\begin{figure*}
\includegraphics[scale=0.5,angle=0]{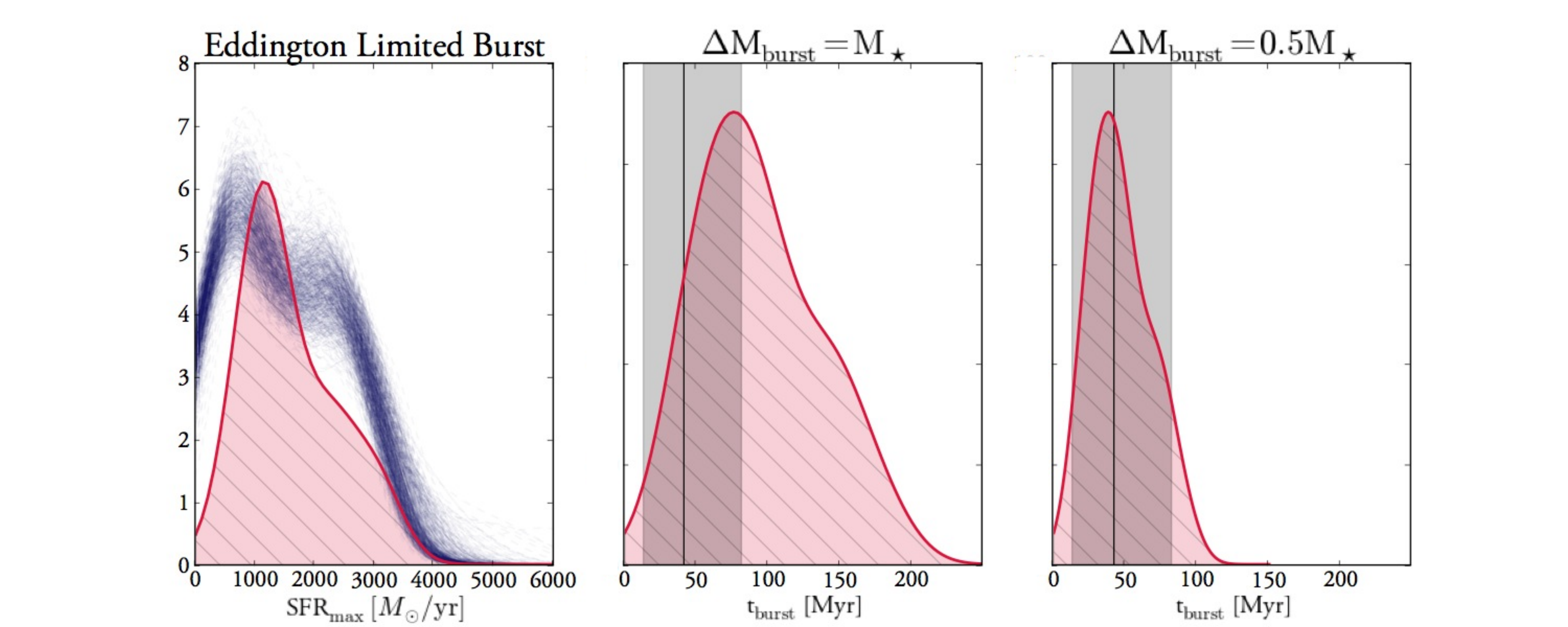}
\caption{Comparison of the SFRs and starburst timescales derived for
  the $z\gtrsim3$ SMGs $z\sim2$ quiescent galaxies.
{Left}: 
The red curve show the probability density distribution (KDE) of SFRs of the
${z\sim2}$ quiescent galaxies during their formation, calculated assuming they formed
in Eddington limited starbursts.
The blue curves show probability distributions for 1000 realizations
of ongoing SFRs in the ${z\gtrsim 3}$ SMGs, estimated from their total
infrared luminosity and associated errors.
The two distributions span the same range, in support of a evolutionary
connection between quiescent galaxies and SMGs. 
{Middle:} Probability density distribution of the duration of the
cQG starbursts, calculated assuming all their observed stellar mass
formed in Eddington limited bursts. The grey area
  indicates the constraints on the duty cycle of the SMG starbursts
  derived from their number density. 
{Right:} Same
  as the middle plot, but assuming that only half of the
  ${z\sim2}$ quiescent galaxies' 
  stellar mass formed in the Eddington limited burst.  The two
  independent measures of $t_{burst}$ are consistent, in agreement with
  ${z \gtrsim3}$ SMGs being progenitors of ${z\sim2}$
  quiescent galaxies}
\label{figure.eddington}
\end{figure*}

Based on the observed line widths and compact spatial extent of
molecular line emitting regions, SMGs are often argued to be maximum
starbursts, i.e to be forming stars at a rate close to the Eddington limit. 
Assuming a spherical symmetric geometry, an isothermal sphere density
structure, a small volume filling factor for molecular gas and a
Chabrier IMF, based on \cite{thompson2005},
\cite{younger2010} approximate this ``maximum star formation rate''  as
\begin{equation}
\label{equation.eddington}
\mathrm{SFR_{MAX}}=480\, \sigma_{400}^2 \, D_\mathrm{kpc} \, \kappa_{100}^{-1}\, \, \, \,\, \, \, \mathrm{[M_{\odot}\, yr^{-1}]}
\end{equation} 
where $\sigma_{400}$ is the line-of-sight gas velocity dispersion in
units of 400~km~s$^{-1}$, $\kappa_{100}$ is the opacity in units of
cm$^2$g$^{-1}$ \citep[usually taken to be
$\approx1$,][]{murray2005,thompson2005}, and $D_\mathrm{kpc}$ is the
characteristic physical scale of the starburst (usually approximated
as the Gaussian FWHM of the line emitting region). 
In Figure \ref{figure.eddington}, the blue curves show probability distributions for a 1000 realizations
of ongoing SFRs in the ${z\gtrsim 3}$ SMGs, estimated from their total
infrared luminosity and associated errors, through Equation \ref{equation.sfr}. The SMGs are forming stars
at high rates $\mathrm{500-3000 M_{\odot}yr^{-1}}$, close to the
Eddington limit.  E.g. \cite[]{younger2010} estimated the maximum
starformation rate of AzTEC4 and AzTEC8 to be in the range $1900-3800 M_{\odot}yr^{-1}$, comparable to the values derived here (see Tab.\ref{table.mdust}).

In the following we investigate if the observed properties of ${z\sim2}$ quiescent
galaxies are consistent with having formed under such conditions.
Assuming that ${z\sim2}$ quiescent galaxies formed in Eddington limited
maximum starbursts, we can estimate the maximum SFR and the duration
of this burst, from the observed size, velocity dispersion and
stellar mass of the quiescent remnants.
In Figure \ref{figure.eddington} the red curve shows the distribution of
$\mathrm{SFR_{MAX}}$ for the sample
of ${z\sim2}$ quiescent galaxies described in Section
\ref{section.quiescent}, calculated from Equation
\ref{equation.eddington}, assuming $\kappa_{100}=1$,
$\mathrm{D_\mathrm{kpc}=2\,r_{e,c}}$ (where $\mathrm{r_{e,c}}$ are the effective
  radii measured for the individual galaxies) and
  $\sigma_{400}=\left<\sigma\right>/400\, \mathrm{km\, s^{-1}}$, where
  $\left<\sigma\right>=363\pm100 \, \mathrm{km \, s^{-1}}$ is the mean velocity dispersion
  measured for ${z\sim2}$ quiescent galaxies in the litterature
  \citep{toft2012}. We use this mean value as measured
  velocity dispersions for the K13 sample are not
  available. 

There is a good general correspondence between 
the $\mathrm{SFR_{MAX}}$ distribution of quiescent ${z\sim2}$
    galaxies,  and the SFR distribution of ${z\gtrsim 3}$ SMGs. The $\mathrm{SFR_{MAX}}$
distribution peaks at higher SFRs than the observed distribution in
${z\gtrsim 3}$ SMGs, indicating that some of the  $z\sim2$
quiescent galaxies may have formed in starbursts with sub-Eddington SFRs.  
Also plotted in Figure
  \ref{figure.eddington} (b) is the duration of this ``maximum starburst''
\begin{equation}
\label{equation.tburst-eddington}
\mathrm{t_{burst}=\Delta M_{\star}/SFR_{MAX}}
\end{equation}  
 assuming a constant star fomation
  rate $\mathrm{SFR}=\mathrm{SFR_{max}}$ during the burst, and that all
  the stellar mass of the $\mathrm{z\sim2}$ quiescent galaxies was created during
  this burst, i.e.\ $\mathrm{\Delta M_{\star} = M_{\star}}$. While consistent within the errors, the mean derived
  timescales for Eddington limited starbursts  is about a factor of
  two longer than the starburst timescale derived from comparing
  comoving number densities. This can be accounted for by changing
  some of the assumptions, e.g., if the SMG starbursts are triggered
  by major mergers, a fraction of the stellar mass must have been formed in
  the progenitor galaxies, prior to the merger.  In Figure
  \ref{figure.eddington} (c) we show that if we assume that only half of the observed stellar mass in ${z\sim2}$
  quiescent galaxies was created in a ${z\gtrsim 3}$ Eddington limited starburst, i.e.\
  $\mathrm{\Delta M_{\star} = 0.5 \,M_{\star}}$, there
  is excellent agreement between the derived timescales, consistent
  with the idea that ${z\gtrsim 3}$ SMGs are the progenitors of
  ${z\sim2}$ quiescent galaxies. Interestingly this is
  consistent with the results of \citet{michalowski2010b} who found
  that on average $\sim 45\%$ of the stellar mass in a sample of
  ${z>4}$ SMGs was formed in their ongoing starbursts. 
If half the stellar mass formed prior to the merger that ignite the
SMG starburst, an implication is that the merger progenitors must have
been gas rich star forming galaxies, in agreement with the high gas
fractions found in high redshift star forming galaxies \citep{tacconi2013}

\subsection{Additional Stellar Mass Growth and Quenching of the $z\gtrsim 3$ SMGs}
\begin{figure}
\includegraphics[scale=0.7,angle=0]{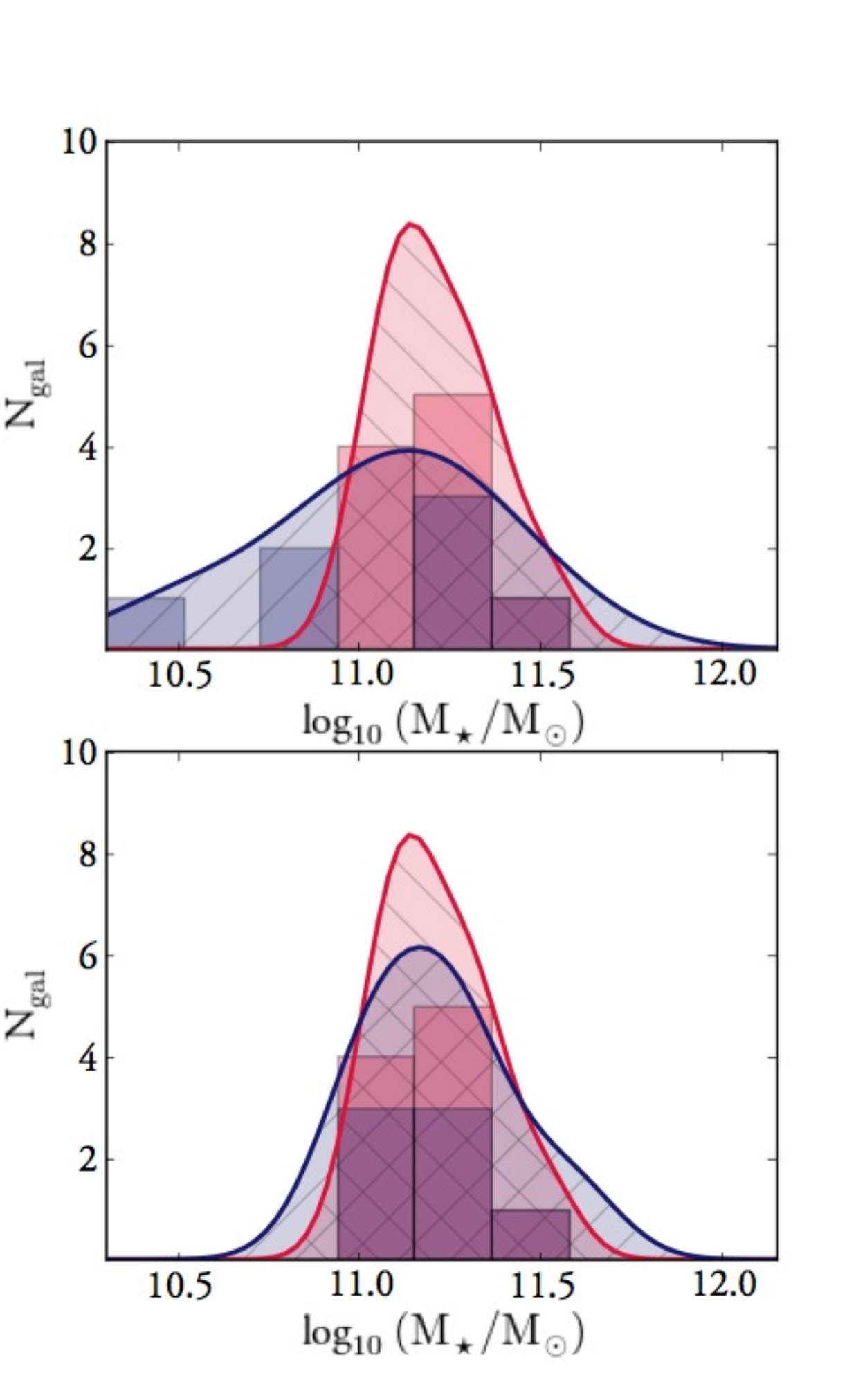}
\caption{ Red histograms show the distribution of stellar masses
 in $\mathrm{log(M/M_{\odot})
  >11}$ quiescent galaxies at ${z\sim2}$. 
 In the top panel the blue histogram show the
 distribution of stellar masses in the ${z\gtrsim 3}$ SMGs. 
In the bottom panel the blue histogram show the final stellar masses of
the ${z\gtrsim 3}$ SMGs assuming that
$10\% \pm5 \%$ of their derived gas mass is turned into stars during
the remainder of the ongoing starburst \cite{hayward2011}. 
}
\label{figure.deltam}
\end{figure}

The similar mass-size distribution of the ${z\gtrsim 3}$ SMGs and ${z\sim2}$
quiescent galaxies is in agreement with what one would expect if the
${z\gtrsim 3}$ SMGs evolve passively into  ${z\sim2}$ quiescent galaxies, after they have been quenched. 
Prior to the quenching however the ongoing starburst will increase the
stellar masses of the galaxies. 
In Figure \ref{figure.deltam} we show that the distribution of stellar masses in the
${z\gtrsim 3}$ SMGs is broader than that of $\mathrm{log(M/M_{\odot})
  >11}$ quiescent galaxies at ${z\sim2}$.
We can estimate the growth of stellar mass in the individual $z\gtrsim 3$ SMGs
from their gas masses, inferred from the FIR SED fits (see Table
\ref{table.mdust}). 
From these we can estimate the final stellar masses of the
${z\gtrsim 3}$ SMGs if we assume a star formation effeciency, i.e the
fraction of gas that is turned into stars during the starburst.
 In the simulations of \cite{hayward2011} the gas fraction decrease
from 45\% to 40\%  in isolated disks and from 17.5\% to 15\% in merging
galaxies,  from the peak of the starburst to when it ends,
corresponding to a decrease in gas mass of 5\% and 15\% during this
time. If we assume that this gas is turned into stars, and that
 we are observing the SMGs at the peak of their starburst, the
models thus indicate that $\sim 10 \pm 5 \%$ of the observed gas mass in
the $z\gtrsim 3$ SMGs will be turned in to stars during the remainder of the burst.
 In Figure \ref{figure.deltam} we compare the final stellar mass distribution of
the ${z\gtrsim 3}$ SMGs with that of quiescent ${z\sim2}$
quiescent galaxies, assuming that $10\%$ of the derived gas mass in
the $z\gtrsim 3$ SMGs are turned into stars before the starbursts are quenched. The two distributions are similar, with a K-S test
statistic of 0.33 and a probability of 67\%, in agreement with a direct
evolutionary link between the two populations.
The mass increase from the time the SMGs are observed up to the end
of the starburst will likely not significantly increase the effective
radii, since the process is highly dissipative, 
resulting in a slight horizontal shift in the $\mathrm{M_{\star}-r_e}$
plane (blue points in Figure \ref{figure.masssize}).
The continued starbursts and subsequent quenching, may also
provide the mechanism needed to transform the observed low-n disk-like surface brightness profiles observed in SMGs to the higher n bulge-like profiles observed in
quiescent galaxies at ${z\sim}2$
\citep{wuyts2011,szomoru2011,bell2012}. Most of the stellar mass will
be added in the nuclear regions of the SMGs, which is likely highly
obscured by dust. Once the quenching sets in and most of the dust
is destroyed or blown away, a more centrally concentrated
surface brightness distribution could be revealed.
Note that if, as assumed here, only 10\% of the large derived gas masses in the $z\gtrsim 3$
SMGs is turned into stars during the remainder of the burst, the following
quenching mechanism must be highly effecient at heating or expelling
the substantial amounts of leftover gas. A possible mechanism for
expelling the gas is through outflows, driven by strong winds
associsted with the maximum starbursts. Tentative evidence for such
outflows have recently been observed in the 163{\micron} OH line profile in
an SMG at z=6.3 \citep{riechers2013}.
We stress that the large systematic uncertainties on the derived
stellar masses for the SMGs could
potentially influence our conclusions.

\section{Summary and Discussion}
\label{section.discussion}

\subsection{The Link between $z\gtrsim 3$ SMGs and $z\sim2$ Compact, Quiescent Galaxies }
In this paper we presented evidence for a direct
evolutionary connection between two of the most extreme galaxy types
in the universe, the highest redshift (${z\gtrsim 3}$) SMGs which host some of
the most intense starbursts known, and quiescent galaxies at ${z\sim2}$ which host
the densest conglomerations of stellar mass known.
The comparison was motivated by the recent discovery of a significant
population of SMG at $3<z<6$ and high resolution imaging and spectroscopic studies of 
${z\sim2}$ quiescent galaxies which show that the majority of their
stars likely formed in massive nuclear, possibly dust eshrouded starbursts in
this redshift range.
From a unique flux-limited statistical sample of
${z\gtrsim 3}$ SMGs in the COSMOS field, we have put robust constraints on
their co-moving number density, which we then put in context of the comoving
number densities of quiescent galaxies of similar mass at ${z\sim2}$.
If ${z\gtrsim 3}$ SMGs are progenitors of ${z\sim2}$ quiescent
galaxies, then our data implies that the SMG duty cycle must be
$\mathrm{t_{burst}=42_{-29}^{+40}Myr}$, where the errorbars include our best estimates of the effects of cosmic
variance, photometric redshift errors and incompleteness.
This timescale is independent from, but in good agreement with estimates based on SMG  gas depletion
timescales $\mathrm{t_{burst}\sim 40-200\, \mathrm{Myr}}$, estimates
from hydrodynamical merger simulations
$\mathrm{t_{burst}\sim50\, \mathrm{Myr}}$ \citep[e.g.][]{mihos1996,cox2008}, and estimates based on the  time, compact starburst galaxies spend
above the main sequence of star formation $\mathrm{t_{burst}<70}$~Myr
\citep{wuyts2011}.
Importantly, as our estimate of the SMG starburst timescale is based
only on number density arguments, it is relatively independent on assumptions
of the underlying stellar inital mass function (IMF), which is a large
potential systematic uncertainty, e.g. in depletion timescale estimates. 

Based on stellar masses derived from UV-MIR photometry and sizes
derived from deep NIR imaging, we have shown that the mass-size
distribution of the ${z\gtrsim 3}$ galaxies is remarkably similar to that observed for 
compact quiescent massive galaxies at ${z\sim2}$, with similar mean internal
stellar mass surface densities  $\mathrm{\left<log(\Sigma)\right > \sim9.8 M_{\odot}{kpc}^{-2}}$.
The surface brightness distributions of the ${z\gtrsim 3}$ SMGs
are best  fit by Sersic models with low Sersic
n parameters, typical of local star forming disk galaxies or mergers, and the
majority show multiple components or irregularities indicative of
ongoing merging and/or clumpy structure. 

Many similarities between ${z\sim2}$ quiescent galaxies and
SMGs exist: they have similar stellar masses, characteristic internal velocities, dynamical
masses, sizes, correlation lengths etc. 
Millimeter measurements of ${z\gtrsim 3}$ SMGs in continuum and CO 
show signatures of merging or rotation
\citep{younger2008,younger2010,riechers2011a,riechers2011b}, with
molecular emission line widths in
the range  $300-700$~km~s$^{-1}$ (with a few outliers), and a mean
$\left< \mathrm{FWHM} \right >=456 \pm 253$~km~s$^{-1}$
\citep{schinnerer2008,daddi2009a,coppin2010,riechers2010,riechers2011a,
  riechers2011b, swinbank2012, walter2012}  similar to stellar
velocity dispersions $\sigma=300-500$~km~s$^{-1}$ measured in ${z\sim2}$ quiescent galaxies.
For example, for AzTEC 3, at $z=5.3$, \cite{riechers2010} measured a CO linewith of $\mathrm{487
km~s^{-1}}$, and a gas depletion timescale of 30~Myr, similar to the SMG
starburst timescale derived here.
At the depth and resolution of the present data, it is impossible to make strong claims
about how many $z\gtrsim 3$ SMGs are in the process of merging. However all the detected galaxies show evidence of close
companions, multiple components or clumpy structure, and have low
derived Sersic indices, consistent with expectations for
merging galaxies. In particular the two galaxies with HST/WFC3 data
appears to be major mergers. 

\begin{figure*}
\includegraphics[scale=0.7,angle=0]{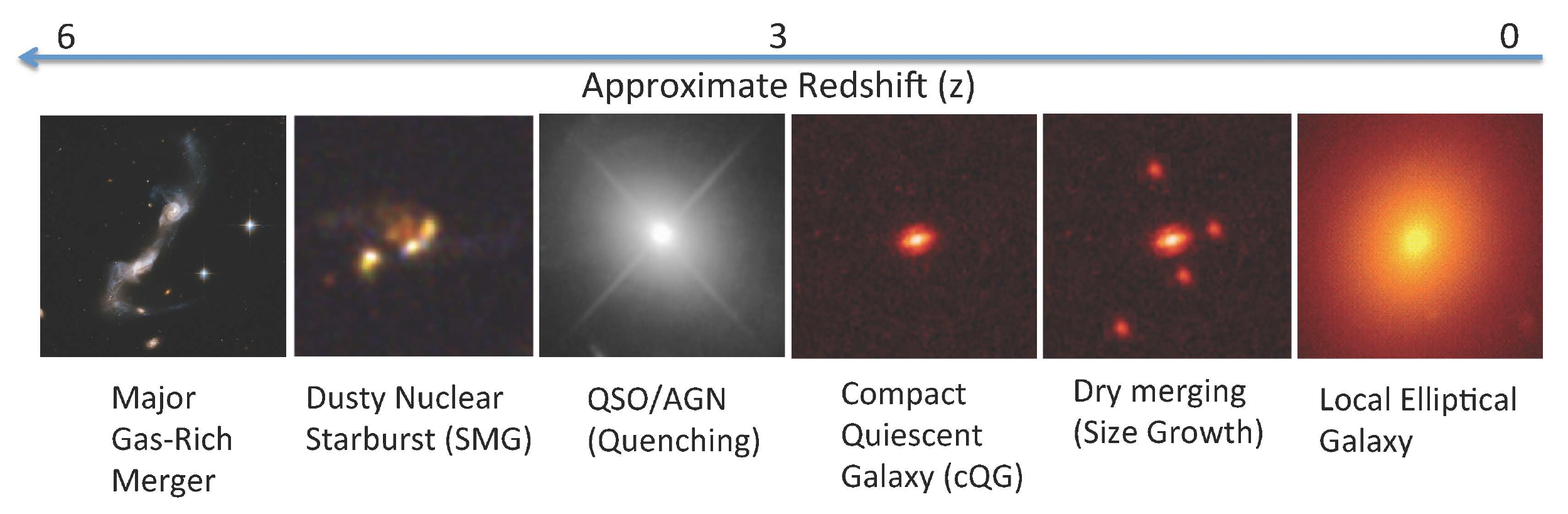}
\caption{Schematic illustration of the formation and evolutionary
sequence for massive galaxies advocated in this paper.}
\label{figure.sequence}
\end{figure*}

The evidence presented in this paper is in support of a direct
evolutionary connection between $\mathrm{z\gtrsim 3}$ SMGs, through compact quiescent
galaxies at ${z\sim2}$, to giant elliptical galaxies in the
local universe.
In this scenario (illustrated in Figure
\ref{figure.sequence}) gas rich, major
mergers in the early universe, trigger nuclear dust enshrouded
starbursts\footnote{The SMG image in the figure is adopted from Targett et al, 2013}, which on average
last $42_{-29}^{+40}$ Myr, followed by star formation quenching,
either due to  gas exhaustion, feedback from the starburst or the
ignition of an AGN, leaving behind compact stellar remnants to evolve
passivly for about a Gyr into the compact quiescent galaxies we
observe at ${z\sim2}$. Over the next 10 Gyr, these then grow
gradually, primarily through minor merging, into local elliptical
galaxies.

\subsection{Connection to Compact Star Forming Galaxies at $2.5<z<3$ }
\label{section.csfg}
\cite{barro2013} found a population of relatively massive (log(M/M$_{\odot}  >
10$) compact star forming galaxies (cSFGs)
at $1.4<z<3$, which show evidence of quenching beginning
to set-in (lower specific star formation rates than typical star forming
galaxies and increased
AGN fractions). Their masses, sizes and number densities (which
increase with decreasing redshift, at the same time the number density of quiescent
galaxies increase), suggest that the highest redshift examples of these
may be progenitors of compact quiescent
${z\sim2}$ galaxies. These galaxies are thus good candidates for transition objects
in the evolutionary sequence suggested here between the ${z\gtrsim 3}$ SMGs and
the ${z\sim2}$ quiescent galaxies. The comoving number density of the
most massive cSFGs ($\mathrm{log(M/M_{\odot} > 10.8}$) at ${2.5<z<3}$ is $\mathrm{\sim 5.4 \pm 2.5
\times 10^{-5}\,Mpc^{-3}}$,  comparable to the number density  
  for 
$z\sim2$ quiescent galaxies. However, the cSFGs are not 
massive enough to be descendants of the brightest ${z\gtrsim 3}$
SMGs, or progenitors of most of the  massive ${z\sim2}$ quiescent galaxies considered here, as none of
the cSFGs have $\mathrm{log(M/M_{\odot}) > 11}$ (Barro, private
communication), but are likely decendents of less intense starbursts
at ${z\gtrsim 3}$ and progenitors of slightly lower mass quiescent ${z=2}$
galaxies.

\subsection{Caveats and Outlook}

 One of the largest uncertainty in the derivables for the ${z\gtrsim 3}$ SMG
 sample are associated with their stellar masses. As extensively
 discusses in \citet{michalowski2012b} stellar masses for SMGs are
 highly dependent on the assumed star formation history, and may
 differ by up to $\pm0.5$~dex given different assumptions and models.
 Dynamical mass considerations may set an upper limit to stellar
 masses, however  the $z\gtrsim 3$ SMGs samples with available
 dynamical mass estimates are still sparse, as well as subject of
 their own biases. 

The sample of ${z\gtrsim 3}$ SMGs is still small, and only partially spectroscopically
confirmed. Future, larger and deeper mm surveys, over multiple fields, will allow for better constraints on the evolution of the
co-moving number density of starburst galaxies, to the highest
redshifts, and to study the effects of cosmic variance.  
This will allow for more detailed tests and modeling of the proposed
scenario in different redshift and mass bins, rather that in the single mass
bin and two redshift ranges as possible with the present data. 
E.g., the proposed scenario implies that the significant population of 
${z\sim2}$ SMGs should evolve into compact, $\sim 1$~Gyr old,
massive post starburst galaxies at ${z\sim1.5}$. Interestingly
  \citet{bezanson2012} recently published a spectroscopic sample
  of galaxies with exactly these properties. Similarly, if compact
  quiescent galaxies at $z\gtrsim3$ are found in the future, the
  properties of these should match those of the highest redshift
  ${z>5}$ SMGs. 
With deeper
data it will also be possible to push to lower star formation rates,
and not only consider the most extreme starbursts. This will likely
provide a way of fitting the $2.5<z<3$ cSFG discussed in Section \ref{section.csfg} into the
evolutionary picture.

Cosmological surface brightnes dimming and the large amounts (and unknown distribution) of dust in SMGs make
them extremely faint in the rest-frame UV and optical, and likely
bias the sizes measured, even in very deep NIR imaging data. 
However, we do note that one of the galaxy in our sample (AzTEC1), has
been resolved in high resolution submillimeter imaging \citep{younger2008}, with a derived
extend of $0.1-0.2\arcsec$, corresponding to physical size of $1.3-2.7$
kpc, consistent with the constraints on the effective radius we
measure from the UltraVISTA data ($\mathrm{r_e<2.6 kpc}$, see Table \ref{table.sample}).   
ALMA will greatly improve estimates of the sizes of high
redshift SMGs, through high resolution observations of the restframe
FIR dust continuum.
We have argued in this paper that the observed structural properties are
consistent with the SMGs being disks or mergers, but the constraints
are uncertain, due to the relatively low S/N and spatial resolution of
the images, e.g.\ the Sersic n parameters and effective radii could be underestimated,  due to obscuration by dust and cosmological surface
brightness dimming.
With ALMA it will be straightforward to determine redshifts from
molecular lines, and constrain the internal dynamics of the galaxies,
e.g.\ estimate velocity dispersions, rotational velocities and search for evidence of merging.
This will provide powerful diagnostics to help map the transformation of the
 most massive galaxies in the universe from
 enigmatic starburst at cosmic dawn to dead remnants, a
few gigayears later. 
 \\
 


ST acknowledges the support of Lundbeck
foundation and is grateful for the hospitality and support of the
Institute for Astronomy, University of Hawaii, during the visit where
this work was initiated.
The research leading to these results has received 
funding from the European Union's Seventh Framework programme under
grant agreement 229517. KS gratefully acknowledges support from Swiss
National Science Foundation Grant PP00P2\_138979/1. MJM acknowledges the support of the FWO-Vlaanderen and the Science and Technology Facilities Council.
K.S. acknowledges support from the National Radio Astronomy
Observatory, which is a facility of the National Science Foundation
operated under a cooperative agreement by Associated Universities,
Inc. J.S acknowledges support through NSF ATI grants 1020981 and 1106284. 
We thank M. Barro for sharing additional
information about his compact star forming galaxies. 
We thank D. Watson and J. Hjorth for helpful discussions.
The Dark Cosmology Centre is funded by the Danish National Research Foundation.


\end{document}